\documentclass[a4paper,10pt,twoside]{cpc-hepnp}

\usepackage{multicol}
\usepackage{graphicx}
\usepackage{booktabs}
\usepackage{amssymb,bm,mathrsfs,bbm,amscd}
\usepackage[tbtags]{amsmath}
\usepackage{lastpage}

\begin{document}

\fancyhead[c]{\small Submitted to 'Chinese Physics C'} \fancyfoot[C]{\small xxx-\thepage}

\footnotetext[0]{Received xxx}

\title{The study and design of RF coupler for Chinese ADS HWR Superconducting Cavity}

\author{%
      MENG Fan-Bo(孟繁博)$^{1,2;1)}$\email{mengfb@mail.ihep.ac.cn}%
\quad CHEN Xu(陈旭)$^{1,2}$%
\quad PAN Wei-Min(潘卫民)$^{1}$
\quad HUANG Tong-Ming(黄彤明)$^{1}$\\
\quad MA Qiang(马强)$^{1}$
\quad LIN Hai-Ying(林海英)$^{1}$
\quad PENG Xiao-Hua(彭晓华)$^{1}$
}
\maketitle

\address{%
1 (Institute of High Energy Physics, CAS, Beijing 100049, China) \\
2 (Graduate University of Chinese Academy of Sciences, Beijing 100049, China)\\
}

\begin{abstract}
  RF power coupler is a key component of the superconducting accelerating system in Chinese ADS proton linac injector I, which is used to transmit 15kW RF power from the power source to the superconducting HWR cavity. According to the requirement of working frequency, power level, transmission capability and cooling condition, the physics design of coupler has been finished, which includes RF structure optimization, thermal simulation, thermal stress analysis and so on. Based on this design, the prototype of HWR coupler has been fabricated, and then has passed the high power test successfully.
\end{abstract}



\footnotetext[0]{\hspace*{-3mm}\raisebox{0.3ex}{$\scriptstyle$}
}%

\begin{multicols}{2}

\section{Introduction}
In Chinese ADS proton linac injector I, HWR superconducting cavity is chosen as the main accelerating structure. In this case, a type of specified coupler should be de developed to meet the requirements of 162.5MHz HWR cavity. While the beam load is reaching 10 mA, the cavity needs to deliver about 15kW RF power to the beam, which means the coupler should have the capability to transmit 15kW CW RF power in travelling wave. Based on the design idea of IFMIF coupler~\cite{lab1} and the successful experience of BEPCII coupler~\cite{lab2}, the structure of coupler’s each part has been determined for further optimization. The parameters of HWR coupler are shown in Table~\ref{tab1}.

\begin{center}
\tabcaption{ \label{tab1}  Parameters of HWR cavity}
\footnotesize
\begin{tabular*}{80mm}{c@{\extracolsep{\fill}}ccc}
  \toprule
  Frequency & 162.5MHz \\
  Impedance & 50ohm \\
  Structure type & Coaxial coupler \\
  Window type & Single, disk ceramic \\
  Coupling method & Antenna coupling \\
  Power level & CW, TW, 15kW \\
  \bottomrule
\end{tabular*}
\end{center}

\section{RF optimization of coupler}

\subsection{Resonant mode in the window}

After the main structure of coupler has been determined, the RF simulation should be done first. The RF structure of coupler is shown in Fig.~\ref{fig1}. As seen from the picture, the coupler is consist of three parts, which are T-box, ceramic window with choke structure and coaxial power transmission line from top to bottom. The window with choke is a local resonant structure~\cite{lab3}. If there are resonant modes near the working frequency 162.5MHz, it will influence the power transmission, and form high local E field which can further cause multipacting or arc.

\begin{center}
\includegraphics[width=5cm]{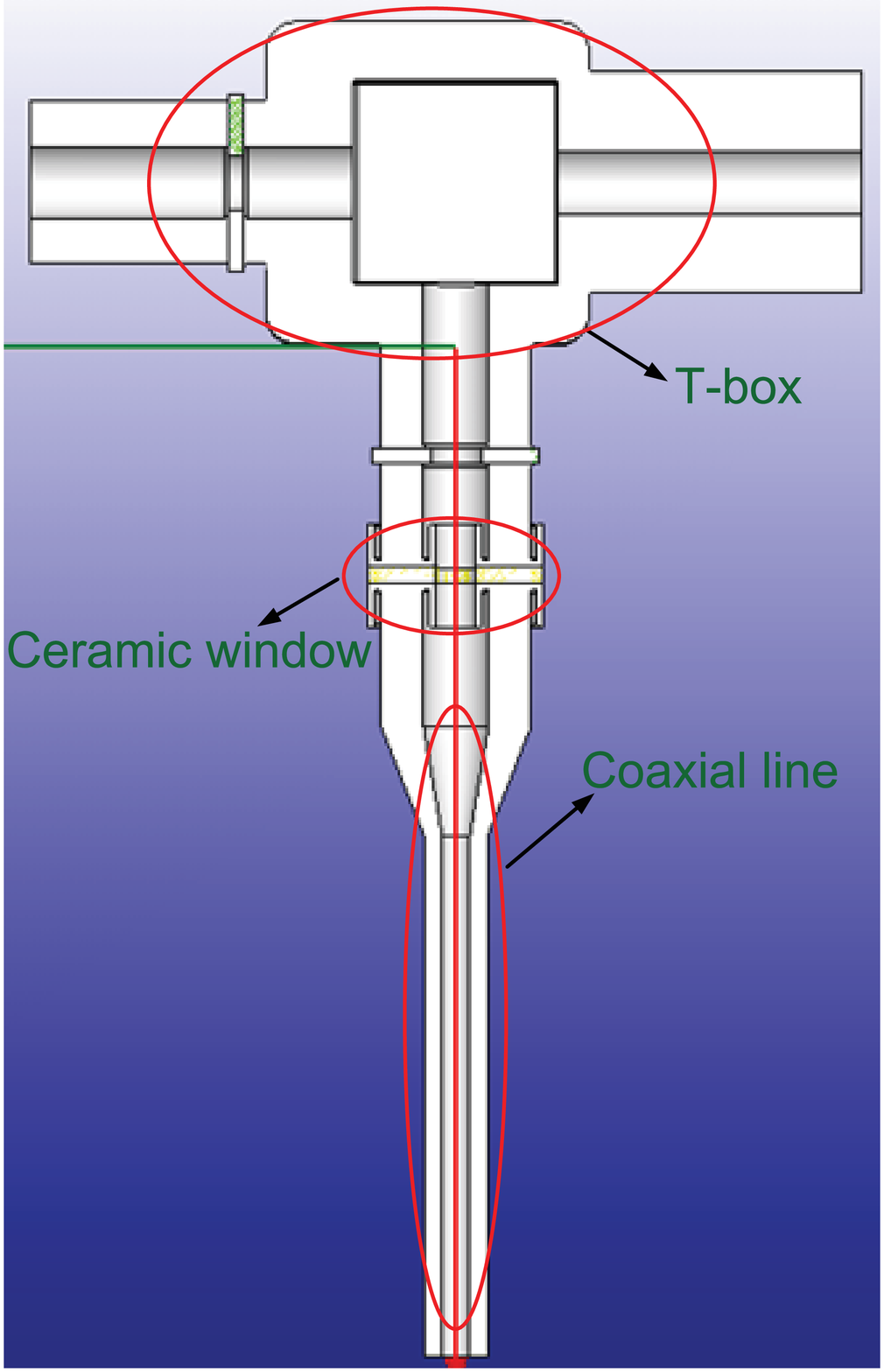}
\figcaption{\label{fig1} RF structure of HWR coupler, consist of T-box, ceramic window and coaxial line }
\end{center}

\noindent 
After the eigen simulation of window structure with two kinds of boundary condition (H boundary and E boundary), the first resonant mode is found at 1.16GHz. This mode is far from the working frequency to make sure the window can work in a safe condition.

\subsection{S-parameter optimization}
Since the window with choke is not a standard coaxial structure, the size of T-box should be chosen seriously to guarantee the impendence matching for the whole coupler. As shown in picture, the left side of T-box is used to connect to the RF power source through the standard Φ105 coaxial line. On the right side of T-box, there is a short-circuit face which can be moved to adjust the mismatching. During the simulation, the size of outer box and inner box, the radius of chamfer and the position of short-circuit face are sensitive for the S-parameter. Finally a good S parameter result has been found. As shown in Fig.~\ref{fig2}, the S11=-40.5dB at central frequency 162.5MHz, and the bandwidth is about 20MHz for S11<-20dB.

\begin{center}
\includegraphics[width=8cm]{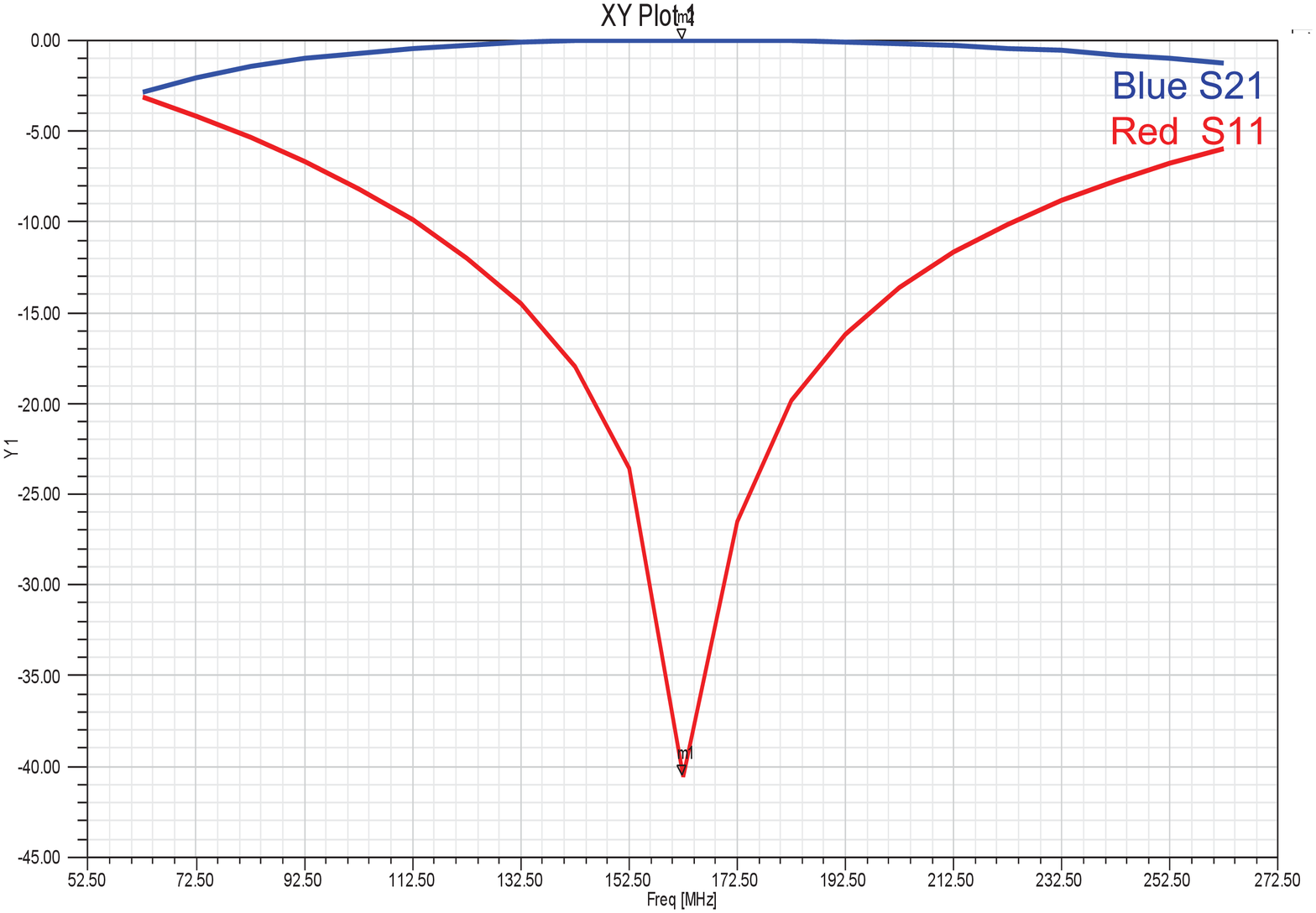}
\figcaption{\label{fig2} Optimized S-parameter of coupler, S11=-40.5dB @162.5MHz }
\end{center}

\section{Thermal stress analysis of ceramic window}
An important function of coupler is to make the separation between air (power source side) and high vacuum (cavity side), so the ceramic window is a core component. When the RF power pass through the window, the dielectric loss of ceramic and Joule loss on the surface of copper will cause temperature rise. As a result, it will change the window’s shape and generate considerable thermal stress~\cite{lab4}. In our calculation, there are two kinds of cooling condition adopted for comparison: one is 25℃ water cooling in inner conductor; the other one is no cooling in inner conductor. In both of the two conditions, the outer wall of the window can transfer heat with the room temperature environment through natural-convection method.\\
In the thermal simulation, the RF power is selected as 15kW equals to the maximum design value. From the temperature distribution results, it can be found that without water cooling the maximum temperature is 321K located at inner conductor, but with water cooling the maximum temperature is just 300K located at outer conductor. Through the comparison, the water cooling is a good method to control the temperature rise of window. Based on the thermal simulation, the thermal stress of widow structure can be further calculated. As shown in Fig.~\ref{fig3}, the maximum thermal stress is 55.1MPa without cooling, and the maximum thermal stress with cooling is only 5.23MPa. Due to the cooling, the thermal stress has been reduced by one order. Since the mechanical strength of ceramic is about 200$\sim$300MPa, this analysis indicates that 15kW is a safe power level for our coupler in real operation.

\begin{center}
\includegraphics[width=8cm]{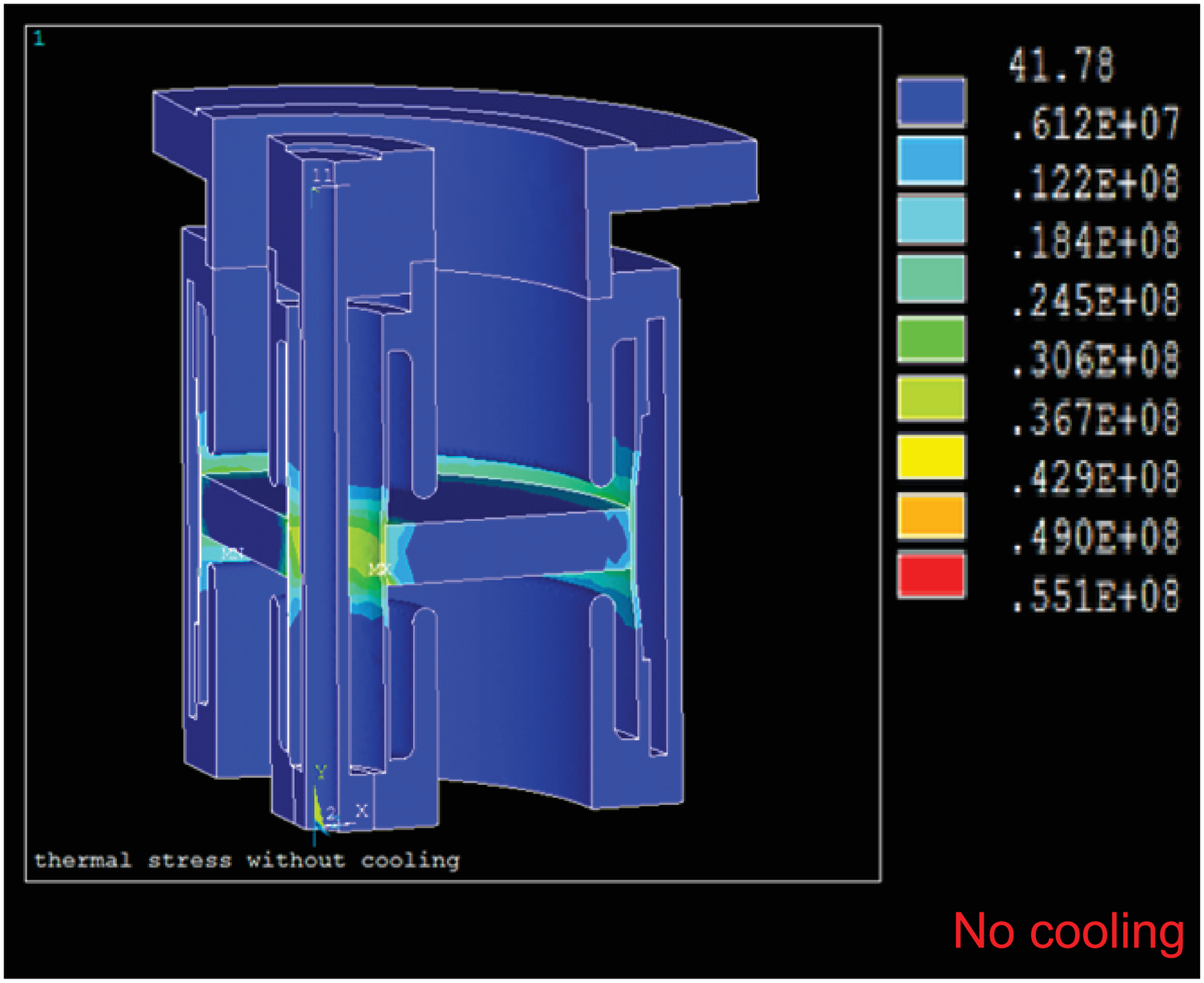}
\\a. thermal stress without cooling
\end{center}

\begin{center}
\includegraphics[width=8cm]{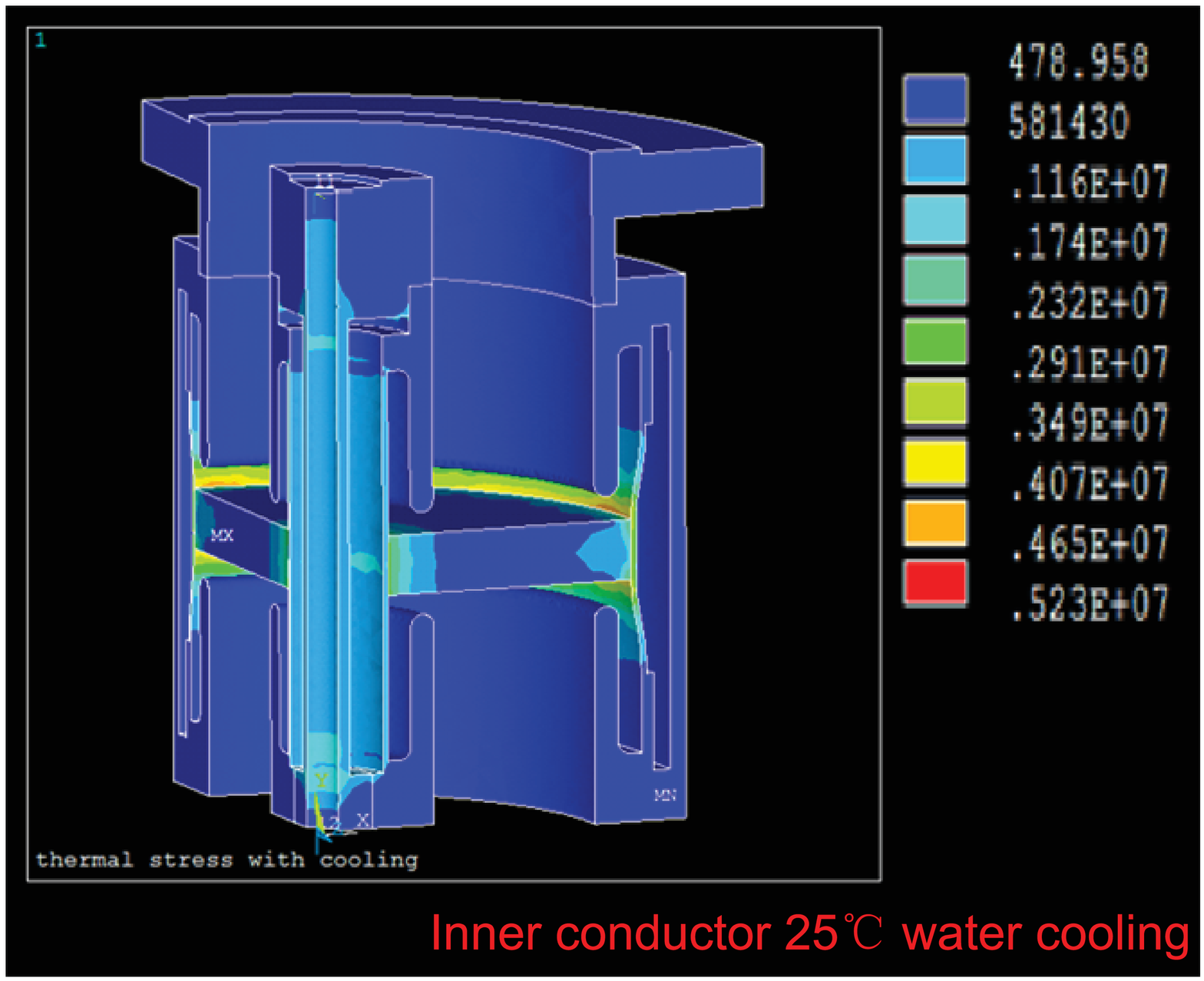}
\\b. thermal stress with water cooling
\figcaption{\label{fig3} Thermal stress distribution of window }
\end{center}

\section{Cooling type of outer conductor}
In superconducting RF system, coupler is a component to achieve the thermal transition between room temperature and liquid He temperature. When the coupler is assembled in the cryomodule, the outer conductor of coupler will transfer heat from outside to the cavity which is working at 2-4K. At the same time, if the RF power is on, the Joule heat generated on the metal surface will also transfer to the cavity through outer conductor. To relieve to the load of refrigerator, the heat leakage through outer conductor should be controlled as low as possible. In the cryomodule, the 80K thermal anchor can be chosen as the first cooling step. Besides that, the He gas, which needs to be recycled from the liquid He evaporation, can be used to cool the outer conductor as the second step~\cite{lab5}.\\

\begin{center}
\includegraphics[width=8cm]{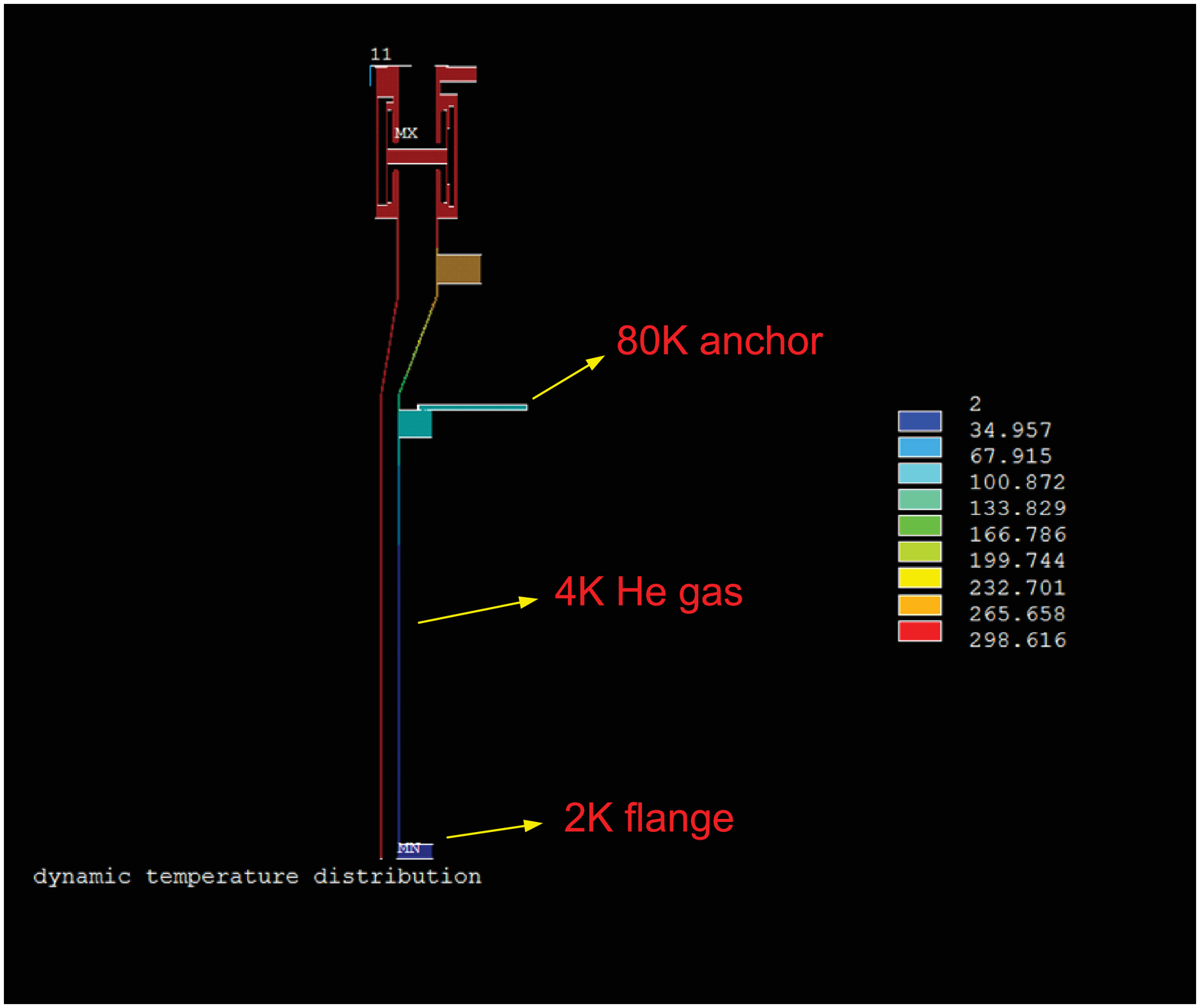}
\figcaption{\label{fig4} Cooling condition and dynamic temperature distribution }
\end{center}

At the beginning of design, the working temperature of HWR cavity is not determined between 2K or 4K. As a conservative estimate, the 2K cavity temperature is selected in the simulation. The cooling condition and the dynamic temperature distribution of coupler are shown in Fig.~\ref{fig4}. When RF power is off, the static heat leakage at 2K is 0.026W. While 15kW RF power passes through the coupler, the dynamic heat leakage at 2K is 0.69W. These values are adequate for the cryomodule requirements. In this case, the combination of 80K thermal anchor and 4K cooling He gas is adopted as our outer conductor’s cooling method.

\begin{center}
\includegraphics[width=6cm]{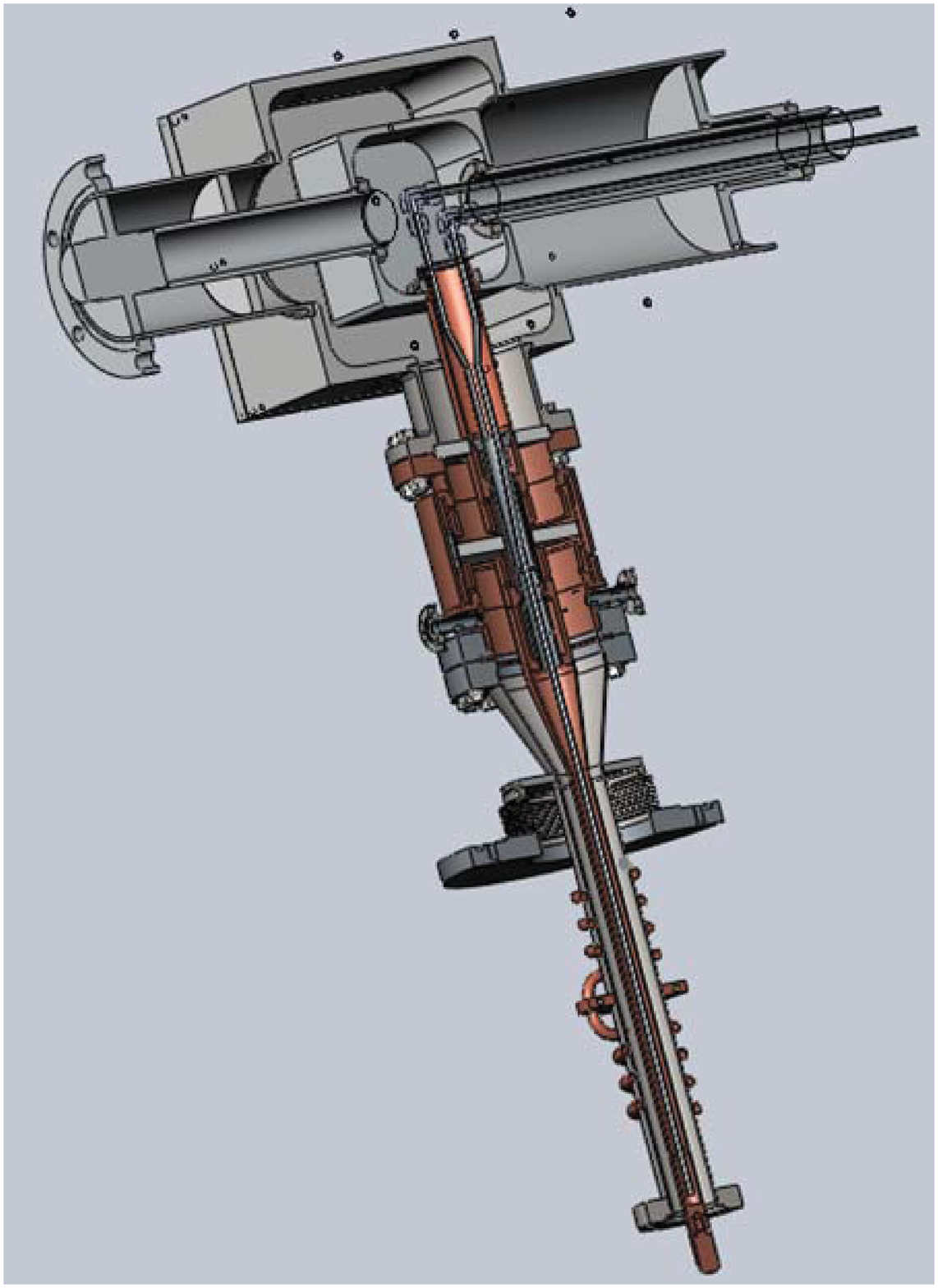}
\figcaption{\label{fig5} Machine drawing of HWR coupler }
\end{center}

\section{Summary}
After the RF, thermal and mechanical design are accomplished, the prototype of coupler has been fabricated (Fig.~\ref{fig5}). Soon later, the high power test has been proceeded to investigate the performance of coupler. During the test, by adjusting the position of short-circuit face, the VSWR of system can be reduced to 1.16, which shows an agreement with our design. Through several days power conditioning, the RF power has passed 15kW and finally reached to 20kW. Due to the limitation of power source, the experiment could not be continued in the higher power level. With the help of water cooling in inner conductor, the temperature of window is maintained around 30$^{\circ}$C all the time. The successful test of prototype means the physics design of coupler is reasonable for the real operation. Since there is still some problem of out-gassing and multipacting at low power in the test, some modification of our design should be done in the future.

\end{multicols}

\vspace{1mm}

\begin{multicols}{2}

\end{multicols}

\vspace{-1mm}
\centerline{\rule{80mm}{0.1pt}}
\vspace{2mm}

\begin{multicols}{2}

\end{multicols}

\clearpage

\end{document}